# Graphene on ferromagnetic surfaces and its functionalization with water and ammonia

S. Böttcher,[1,2] M. Weser,[1] Yu. S. Dedkov,[1,*] K. Horn,[1] E. N. Voloshina,[2,*] and B. Paulus[2]

[1]*Fritz-Haber-Institut der Max-Planck-Gesellschaft, 14195 Berlin, Germany*
[2]*Institut für Chemie und Biochemie, Freie Universität Berlin, 14195 Berlin, Germany*

[*]Corresponding authors:
Tel.: +49-(0)30-84135628
Fax: +49-(0)30-84135603
E-mail: dedkov@fhi-berlin.mpg.de;

Tel.: +49-(0)30- 838 53757
Fax: +49-(0)30- 838 54792
E-mail: elena.voloshina@fu-berlin.de

**Abstract**

Here we present an angle-resolved photoelectron spectroscopy (ARPES), x-ray absorption spectroscopy (XAS), and density-functional theory (DFT) investigations of water and ammonia adsorption on graphene/Ni(111). Our results on graphene/Ni(111) reveal the existence of interface states, originating from the strong hybridization of the graphene π and spin-polarized Ni 3$d$ valence band states. ARPES and XAS data of the $H_2O$ ($NH_3$)/graphene/Ni(111) system give an information about the kind of interaction between adsorbed molecules and graphene on Ni(111). The presented experimental data are compared with the results obtained in the framework of the DFT approach.

Graphene is a single layer of carbon atoms arranged in a honeycomb lattice with two crystallographically equivalent atoms (C1 and C2) in its primitive unit cell [1, 2]. The $sp^2$ hybridization between one 2$s$ orbital and two 2$p$ orbitals leads to a trigonal planar structure with a formation of strong σ bonds between carbon atoms that are separated by 1.42 Å. These bands have a filled shell and, hence, form a deep valence band. The unaffected 2$p_z$ orbital, which is perpendicular to the planar structure of the graphene layer, can bind covalently with neighboring carbon atoms, leading to the formation of a π band. Since each 2$p_z$ orbital has one extra electron, the π band is half filled. The π and π* bands touch in a single point at the Fermi energy ($E_F$) at the corner of the hexagonal graphene's Brillouin zone, and close to this so-called Dirac point the bands display a linear dispersion and form perfect Dirac cones. Thus, undoped graphene is a semimetal ("zero-gap semiconductor"). The linear dispersion of the bands results in quasiparticles with zero mass, so-called Dirac fermions.

The unique "zero-gap" electronic structure of graphene, however, leads to some limitations for application of this material in real electronic devices. In order, for example, to prepare a practical transistor, one has to have a graphene layer where energy band gap is induced via application of electric field or via modification of its electronic structure by means of functionalization. There are several ways of the modification of the electronic structure of graphene with the aim of gap formation [3]. Among them are (i) incorporation in its structure of nitrogen and/or boron or transition-metal atoms; (ii) using different substrates that modify the electronic structure; (iii) intercalation of different materials underneath graphene grown on different substrates; (iv) deposition of atoms or molecules on top; etc.

Here we present an attempt to modify the electronic structure of graphene via contact of this material with metal (ferromagnetic Ni substrate) and via adsorption of polar molecules ($H_2O$, $NH_3$) on top of the graphene/metal system. These studies of water and ammonia adsorption on graphene/Ni(111) were performed via combination of experimental [angle-resolved photoelectron spectroscopy (ARPES), x-ray absorption spectroscopy (XAS)] and theoretical methods [density-functional theory (DFT) calculations]. XAS and ARPES studies of graphene/Ni(111) reveal the existence of the interface states, originating from the strong hybridization of the graphene π and Ni 3$d$ valence band states with partial charge transfer of the spin-polarized electrons on the graphene π* unoccupied states. This leads to the appearance of induced magnetism in the carbon atoms of the graphene layer as confirmed by x-ray magnetic circular dischroism (XMCD). ARPES and XAS data of the $H_2O$,$NH_3$/graphene/Ni(111) systems permit to discriminate between different strengths of interaction (physisorption or chemisorption), which appear between adsorbed molecules and graphene on Ni(111). DFT calculations were used to model different geometries of the adsorbed molecules on top of graphene/Ni(111), and electronic structure calculations were performed for them.

These and previous theoretical studies are compared with our present experimental results.

The ARPES and XAS studies were performed at the BESSY UE56/2-PGM-1 and UE56/2-PGM-2 beam-lines and MAX-lab D1011 beam-line, respectively. An ordered graphene overlayers were prepared on Ni(111) via thermal decomposition of propene ($C_3H_6$) according to the recipe described elsewhere [4, 5, 6]. The quality, homogeneity, and cleanliness of the graphene/Ni(111) system were verified by means of low-energy electron diffraction (LEED) and core-level as well as valence-band photoemission. Water and ammonia were deposited at the partial pressure of $p = 5\times10^{-8}$ mbar on the surface of graphene/Ni(111) at 80 K and sample was kept at this temperature during spectroscopic measurements. XAS and XMCD spectra were collected at both Ni $L_{2,3}$ and C $K$ absorption edges in partial and total electron yield modes (PEY and TEY, respectively) with an energy resolution of 80 meV. ARPES experiments were performed on experimental station allowing to obtain 3D data sets of the photoemission intensity $I(E_{kin},k_x,k_y)$, where $E_{kin}$ is the kinetic energy of emitted photoelectrons and $k_x$, $k_y$ are the two orthogonal components of the wave-vector of electron. The energy/angular resolution in ARPES measurements was set to 80 meV/0.2°. The base pressure during all measurements was below $7\times10^{-11}$ mbar.

In our DFT studies, the electronic and structural properties of the graphene-substrate system have been obtained using generalized gradient approximation, namely the Perdew–Burke–Ernzerhof (PBE) functional, to the exchange correlation potential. For solving the resulting Kohn-Sham equation we used the Vienna Ab Initio Simulation Package (VASP) with the projector augmented wave basis sets [7]. The $k$ -meshes for sampling the supercell Brillouin zone are chosen to be as dense as $24 \times 24$, when folded up to the simple graphene unit cell. Plane wave cutoff was set to 875 eV.

As was previously found [8] and confirmed in our present calculations the most energetically favorable is the *top-fcc* arrangement of carbon atoms on Ni(111) (see Fig. 1). For this structure, several high symmetry adsorption positions for molecules are possible. They are: T- *on-top*, B- *on-bond,* and C- *center* and are marked by the corresponding capital letters in Fig. 1. The are up to 42 and 16 possible configurations of $H_2O$ and $NH_3$, respectively, on top of graphene/Ni(111), but in our calculations we restrict ourselves to 6 arrangements where molecules are placed in the high symmetry positions (T, B, C) with hydrogen atoms pointing upwards (UP) or downwards (DOWN). Two examples of possible absorption geometries are shown for $H_2O$ (C-DOWN, hydrogen atoms are pointed to the direction of C-C bond) and $NH_3$ (T-UP, hydrogen atoms are pointed to the direction of the neighboring C atoms) in Fig. 1. In our experiments we studied molecular layers of adsorbate with the thicknesses of approximately 1/3-1/2 of the molecular layer, ML (corresponding to the dense packing of molecules, when one molecule is placed in every carbon ring). For simplicity, in our calculations the concentration of adsorbed molecules was chosen as 1/3 ML that corresponds to the $(\sqrt{3}\times\sqrt{3})R30°$ overstructure with respect to the unit cell of graphene (shown in Fig. 1 as dashed- and solid-line rhombus, respectively).

In order to study the growth modes of water or ammonia, the time sequences of the photoemission maps around the Γ point of the Brillouin zone (sampling angle of ±10° with respect to the normal emission) were recorder. The extracted photoemission intensity map showing the modification of the valence band at the Γ point of the graphene/Ni(111) system upon adsorption of water molecules ($t$ is the deposition time) is shown in Fig. 2 (central panel). Photoemission intensity profiles for several time-points demonstrating the main photoemission features of spectra [Ni $3d$ states, graphene π states, and water induced states (I and II)] as well as intensity profiles as a function of water deposition time ($t$) taken at particular binding energies (red solid line, blue solid circles, and green open squares show intensity profiles at 7 eV, 8.3 eV, and 10 eV of the binding energy, correspondingly) are shown in the upper and right panels, respectively. The behavior of the water-related photoemission features, I and II, allows to conclude that islands-type growth of water on graphene/Ni(111) takes the place: (i) these features start to grow simultaneously at $t = 130$ sec, but slopes of the intensities growth are different; (ii) after $t = 170$ sec the intensity of feature I decreases via the exponential law and there is a small plateau for the feature II (1st ML is complete); (iii) at $t = 230$ sec, when the thickness of deposited water is more than 2ML, probably, the structural phase transition takes the place – formation of ice. Due to the fact that ice is an insulator, the rapid decrease of the photoemission intensities of the Ni-related features and the shift of some states to higher binding energies can be explained by the formation of an insulating thin film of ice on top of the graphene/Ni(111) system. The delay in starting of the growth of the water-related photoemission features is somewhat puzzling (130 sec until the first water-related signal appears in the spectra), but could be related to the fact that some clustering centers on the graphene/Ni(111) surface are necessary in order to water starts to grow. As soon as enough number of such centers is formed, the process of growth is accelerated. The general trend in the observation of the ammonia-related photoemission features in the similar experiments is the same. In further XAS and ARPES experiments we chose the thickness of water and ammonia layers to be 1/3-1/2 ML (see discussion about the structure earlier in the text).

The effect of the possible orbital mixing of the valence band states of the graphene layer on Ni(111) and orbitals of water and ammonia molecules was studied by XAS (Fig. 3). This figure shows the angular dependence of the C $K$-edge XAS spectra of (a) graphene/Ni(111) and this system after adsorption of 1/2 of the molecular layer of (b) $H_2O$ and (c) $NH_3$, respectively.

The XAS spectra of the clean graphene/Ni(111) system [Fig. 3(a)] were analyzed in details in Refs. [6,9]. According to the theoretical calculations for this system, the first sharp feature in the XAS spectrum at 285.5 eV of photon energy is due to the transition of the electron from the C $1s$ core level into the interface state above the Fermi level (around the $K$ point in the hexagonal Brillouin zone) which originates from the C $p_z$ – Ni $3d$ hybridization and corresponds to the antibonding orbital between a carbon atom C-*top* and an interface Ni atom. The second peak in the XAS spectrum at 287.1 eV of photon energy is due to the dipole transition of an electron from the C $1s$ core level into the interface state above the Fermi level (around the $M$-point in the hexagonal Brillouin zone) which originates from C $p_z$ – Ni $p_x,p_y,3d$ hybridization and corresponds to a bonding orbital between C-*top* and C-*fcc* atoms, involving a Ni interface atom. As was found in the experiment the observed hybridization leads to the orbital mixing of the valence band states of graphene and Ni and to the appearance of the effective magnetic moment of carbon atoms in the graphene layer. This moment was detected in the recent XMCD measurements of this system [6], which allow estimating the spin-magnetic moment of carbon in the range 0.05-0.1 $\mu_B$ per atom.

The XAS spectra of the $H_2O$/graphene/Ni(111) and the $NH_3$/graphene/Ni(111) systems measured at the C $K$ absorption threshold are shown in Fig. 3 (b) and (c), respectively. These results demonstrate the controllable way of the graphene functionalization by water and ammonia and the corresponding adsorbate-induced states in the region of the unoccupied valence band states were detected (Fig. 3: the photon energies region of 280-290 eV corresponds to the C $1s\rightarrow\pi^*$ transitions; the photon energies region of 290-320 eV corresponds to the C $1s\rightarrow\sigma^*$ transitions). Here we would like to emphasize that the presented x-ray absorption spectra were recorded at the C $K$ absorption edge and reflect (to some extent) the partial density of states of the carbon atoms in the system [10] and they are clearly demonstrate the appearance of the orbital hybridization of the graphene- and water- and ammonia- related states, respectively. The absence of the strong angular variation of the water- and ammonia-induced XAS signal might be explained by a statistically uniform distribution of the orientations of $H_2O$ and $NH_3$ molecules on graphene/Ni(111).

The interpretation of the XAS spectra measured after water or ammonia adsorption can be performed on the basis of the peak-assignment, which was presented above. For water adsorbate the new structure in the XAS spectra appears at the photon energy range corresponding to the hybrid state in the electronic structure of graphene/Ni(111) involving both carbon atoms in the unit cell of graphene and interface Ni atom. This leads to the assumption that water molecules are adsorbed either in the *center* or in the *on-bond* position on graphene/Ni(111) (Fig. 1). Ammonia-induced spectral features in the C $K$ XAS spectra are observed in the photon energy range corresponding to the hybrid state which is a result of hybridization of the $p_z$ orbital of the C-*top* atom and the $3d_{z2}$ state of the Ni interface atom. On the basis of this analysis one can conclude that ammonia molecules are placed in the *on-top* position on graphene/Ni(111) with the lone-pair toward carbon atoms and N-H bonds along C-C bond of the graphene layer.

Fig. 4 shows a series of angle-resolved photoemission spectra collected with the photon energy $h\nu$ = 75 eV along the $\Gamma$–$K$ direction of the Brillouin zone for the graphene/Ni(111), $H_2O$/graphene/Ni(111), and $NH_3$/graphene/Ni(111) systems. In all series one can clearly discriminate the dispersions of graphene π- and σ- derived states in the region below 2 eV of the binding energy as well as Ni $3d$- derived states near $E_F$. The binding energy difference of ≈2.4 eV for the π states and ≈1 eV for the σ states in the center of the Brillouin zone (in the $\Gamma$ point) between graphite and graphene on Ni(111) is in good agreement with previously reported experimental and theoretical values [4, 5, 8] and it is explained by the different strength of hybridization for π and σ states with Ni $3d$ states. The effect of hybridization between Ni $3d$ and graphene $\pi$ states can be clearly demonstrated in the region around the $K$ point of the Brillouin zone: (i) one of the Ni $3d$ bands at 1.50 eV changes its binding energy by $\approx$150 meV to larger binding energies when approaching the $K$ point; (ii) a hybridization shoulder is visible in photoemission spectra which disperses from approximately 1.6 eV to the binding energy of the graphene $\pi$ states at the $K$ point. The full analysis of the electronic band structure and magnetic properties of the graphene/Ni(111) system were performed in Ref. [9].

Adsorption of 1/2 ML of water and ammonia molecules on graphene/Ni(111) leads to the appearance of the additional photoemission signal in the spectra at 6.5 eV and 7.3 eV, respectively (Fig. 4). In these spectra these emissions are associated with the $H_2O$-$3a_1$ and $NH_3$-$1e$ states, respectively. As can be clearly seen from the photoemission spectra, the adsorption of $H_2O$ or $NH_3$ on graphene/Ni(111) leaves the electronic structure of graphene $\pi$- and Ni $3d$-states almost intact. This observation can be taken as an indication of the inertness of the graphene layer on Ni(111) as was earlier demonstrated in Ref. [4]. There are only small changes of the electronic structure of graphene/Ni(111) upon adsorption of water or ammonia. The small shift of about 150 meV of the graphene $\pi$ band to the small binding energies is detected at the $\Gamma$ point of the Brillouin zone in both cases. At the $K$ point there is a shift of this band to the higher binding energies of about 50 meV and 70 meV for the water an ammonia adsorption, respectively.

Thus, ARPES and XAS data allow us to refer the interaction between considered molecules and the graphene/Ni(111) system as physisorption. From a theoretical point of view, physisorption can be considered as weak interaction arising due to two types of forces, namely dispersion forces and/or classical electrostatic ones. The dispersion interactions are long-range electron correlation effects, which are not captured in DFT because of the local character of common functionals. Consequently, DFT often fails to describe physisorption correctly. For a correct and consistent treatment of physisorption interaction it is necessary to use high-level wave-function-based post-Hartree–Fock methods like the Møller–Plesset perturbation theory [11] or the coupled-cluster method (CC) [12]. One problem here is that a very accurate treatment, e.g. with the CC method, scales very unfavorably with the number of electrons in the system. In general, this difficulty is avoidable by employing the so-called method of increments, where the correlation energy is written in terms of contributions from localized orbital groups [13]. An alternative approach is an inclusion of the dispersion correction to the total energy obtained with standard DFT approximation explicitly by hand with e.g. DFT-D method, that is atom pair-wise sum over $C_6R^{-6}$ potentials (see, for example Ref. [14]).

Recently, first-principles studies for single $H_2O$ molecule adsorbed on freestanding graphene were performed by O. Leenaerts and co-workers [15]. For comparison reason, the reported interaction energies ($E_{int}$) are listed in Table 1 together with the corresponding equilibrium distances ($d_0$). (Our test VASP-calculations for (3×3) supercell yield similar values). One can observe very low interaction energies and no energetic preference regarding the adsorption site or orientation of the adsorbate. We have repeated the calculations taking into account the dispersion correction as proposed by S. Grimme (DFT-D2 method [16]). The resulting interaction energies are by 4-7 times higher in magnitude, although still physisorption is predicted coincidently with experimental observations. Consequently, the equilibrium distances between $H_2O$ and graphene are significantly shorter. In addition, DOWN orientation is clearly more preferred in this case as compared to the opposite one (*i.e.* UP). Note: the obtained results are in reasonable agreement with the recent CCSD(T) data evaluated for the $H_2O$/graphene system [17]. Thus, for the further consideration of the systems of interest we will use the PBE-D2 approximation.

The results obtained for the ($\sqrt{3}\times\sqrt{3})R30^o$ overstructures of adsorbed molecules on graphene/Ni(111) are presented in Table 2. Due to symmetry breaking by the Ni(111) support, two inequivalent carbon atom in *on-top* positions have to be considered in these cases. The difference between adsorption behavior of water on graphene and graphene/Ni(111) indicates the effect of the substrate underneath of the graphene layer and can be explained by the fact, that in the latter case the electron charge density is shifted to the interface between the graphene layer and the Ni(111) support. At the same time, similarly with the case when free-standing graphene is used as a substrate, for the $H_2O$/graphene/Ni(111) system, DOWN orientation is the energetically most favorable one and the preferable adsorption site is *center* of the carbon ring. This theoretical observation confirms our prediction based on the interpretation of XAS spectra. Note: during these calculations structural optimization of the system was not preformed and only distance between graphene and adsorbate is relaxed. Full optimization of $H_2O$ geometry in the case of C_DOWN configuration leads to $d_0$ = 2.51 Å, that is a deviation of 2 % with respect to the non-relaxed value. The corresponding interaction energy is by 4 % lower, than $E_{int}$ given in Table 2.

For the most stable arrangement of $H_2O$ on top of graphene/Ni(111) the band structure calculations were performed. One finds the $H_2O$-related states at the following binding energies: 3.97 eV, 5.96 eV, 9.85 eV, which satisfactory match the APRPES data.

One can see, when looking at data listed in Table 2, that in the case of ammonia, its interaction energy with the substrate is higher compared to the values obtained for the $H_2O$/graphene/Ni(111) system, that is also in good agreement with the experimental results, where the modification of the XAS C $K$ spectra was observed. Here the UP orientation is preferable for any adsorption position. Although *on-top* (T_C1) adsorption yields the highest interaction energy, one has to be aware that the present calculations cannot give exact answer regarding the energetically most favorable adsorption position since obtained interaction energies are very close to each other (within 3 %). Geometry optimization can make this difference more pronounced, especially when taking into account stronger interaction between ammonia and the considered substrate.

Overall, from a theoretical side we can see good agreement between experimental data and the ones obtained by means of DFT calculations. However, further investigations are required before making the final conclusion regarding the position and orientation of the adsorbate with respect to the substrate under study. Firstly, all possible arrangements of $H_2O$ and $NH_3$ on top of graphene/Ni(111) have to be considered. Optimization of molecular geometry as well as relaxation of interlayer distances within the substrate has to be performed. Furthermore, parameter-free way of accounting for dispersion corrections is preferable. The latter is possible via van der Waals density functional (vdW-DF), developed by Dion *et al* [18].

*In conclusion*, we have studied the modification of the electronic structure of the graphene/Ni(111) system upon adsorption of water and ammonia molecules at low temperature. Adsorption of both types of adsorbates leads to the modifications of the XAS C $K$-edge spectra indicating the orbital mixing of the valence band states of graphene and adsorbates. For the occupied states, the small shifts of the graphene π states were detected in both cases with overall shift of the graphene π states to the lower binding energies reflecting the effect of *p*-doping (with respect to the initial state) after adsorption of water and ammonia on graphene/Ni(111). Analysis of experimental results brings us to the idea of the site-selective adsorption: water is adsorbed either in the center of carbon ring or on the bond between two carbon atoms; ammonia molecules are adsorbed on the carbon atom, which is located above the Ni interface atom. This assumption is supported by the results obtained via DFT calculations.

Acknowledgements

We would like to acknowledge A. Preobrajenski (Max-lab) for the technical assistance during experiment. S.B., M.W., Y.D. acknowledge the financial support by MAX-laboratory (Lund). Y.D. acknowledges the financial support by the German Research Foundation (DFG) under project DE 1679/2-1. E. V. appreciate the support from the German Research Foundation (DFG) through the Collaborative Research Center (SFB) 765 "Multivalency as chemical organisation and action principle: New architectures, functions and applications". We appreciate the support from the HLRN (High Performance Computing Network of Northern Germany) in Berlin.

References:

1. A. K. Geim and K. S. Novoselov, Nature Mater. **6**, 183 (2007).
2. A. K. Geim, Science **325**, 1530 (2009).
3. D. W. Boukhvalov and M. I. Katsnelson, J. Phys.: Condens. Matter **21**, 344205 (2009).
4. Yu. S. Dedkov *et al*., Appl. Phys. Lett. **92**, 052506 (2008).
5. Yu. S. Dedkov *et al*., Phys. Rev. Lett. **100**, 107602 (2008).
6. M. Weser *et al*., Appl. Phys. Lett. **96**, 012504 (2010).
7. G. Kresse and J. Hafner, J. Phys.: Condens. Matter **6**, 8245 (1994).
8. G. Bertoni *et al*., Phys. Rev. B **71**, 075402 (2004).
9. Yu. S. Dedkov and M. Fonin, New J. Phys., accepted (2010).
10. J. Stöhr, *NEXAFS Spectroscopy*, 2nd ed. (Springer-Verlag, June 1996).
11. C. Møller and M.S. Plesset, Phys. Rev. **46**, 618 (1934).
12. J. Cizek, Adv. Chem. Phys. **14**, 35 (1969).
13. B. Paulus and K. Rosciszewski, Int. J. Quantum Chem. **109**, 3055 (2009).
14. S. Grimme, J. Comput. Chem. **25**, 1463 (2004).
15. O. Leenaerts *et al*., Phys. Rev. B **77**, 125416 (2008).
16. S. Grimme, J. Comp. Chem. **27**, 1787 (2006).
17. E. Voloshina *et al.*, unpublished.
18. M. Dion *et al.*, Phys. Rev. Lett. **92**, 246401 (2004).

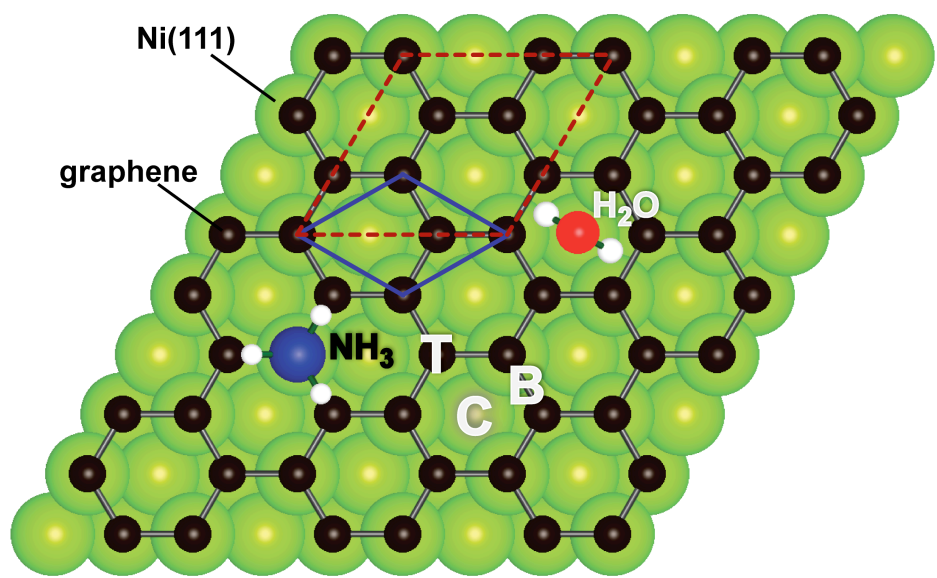

Fig. 1. Geometry of the H₂O,NH₃/graphene/Ni(111) systems studied in this work. Graphene layer is arranged in the *top-fcc* configuration on Ni(111). Adsorbed molecules can be placed in three different highly symmetric adsorption sites: T- *on-top*, B- *on-bond*, C- *center*, with respect to the graphene lattice. Two examples of adsorption are shown: for NH₃ in the *on-top* position with hydrogen atoms directed to the neighboring carbon atoms and for H₂O in the *center* position with hydrogen atoms directed to the C-C bonds.

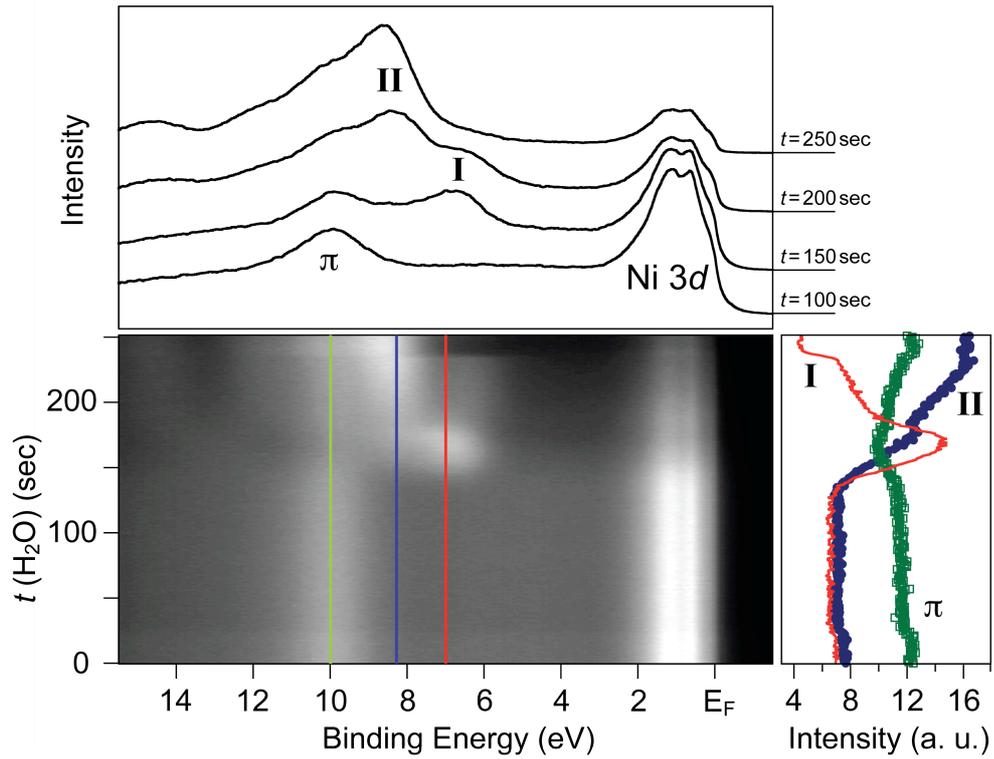

Fig. 2. (Central panel) Photoemission intensity map shows the modification of the valence band of the graphene/Ni(111) system at the Γ point upon adsorption of water molecules (partial water-pressure $p=5\times10^{-8}$ mbar; $t$ is the deposition time). (Upper panel) Photoemission intensity profiles are shown for several time-points demonstrating the main photoemission features: Ni 3$d$ states, graphene π states, and water induced states (I and II). (Right panel) Photoemission intensity profiles as a function of water deposition time ($t$) taken at particular binding energies: red solid line, blue solid circles, and green open squares show intensity profiles at 7 eV, 8.3 eV, and 10 eV of the binding energy.

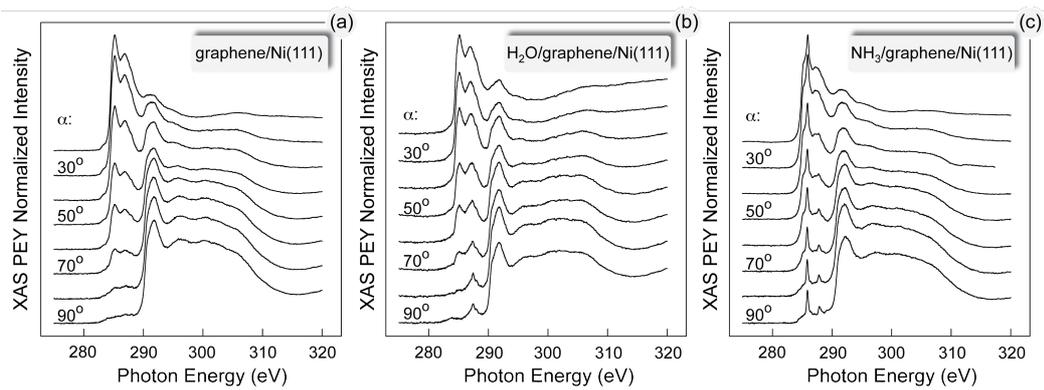

Fig. 3. Angular dependence of the C *K*-edge XAS spectra of (a) graphene/Ni(111) and this system after adsorption of one-half of the molecular layer of (b) $H_2O$ and (c) $NH_3$, respectively.

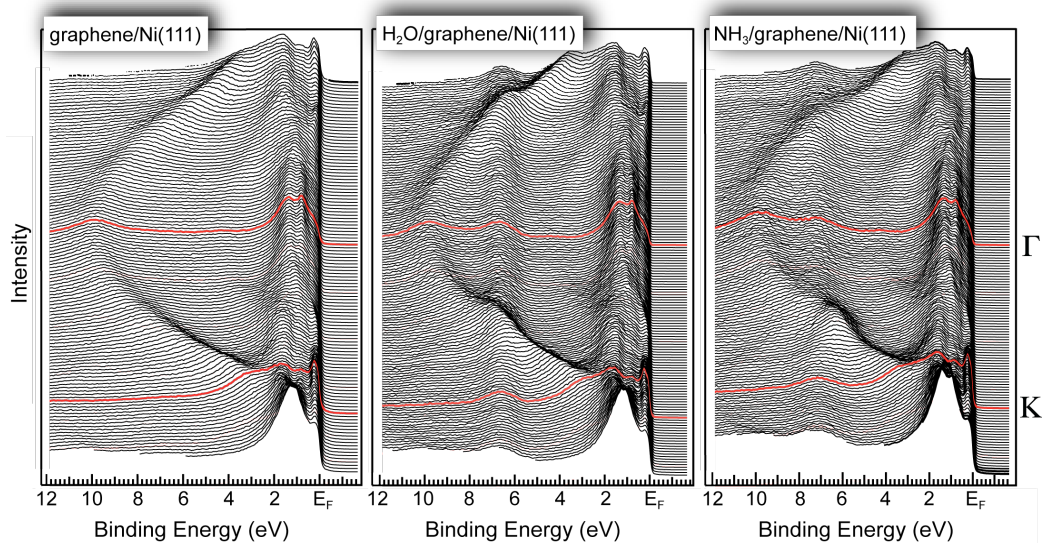

Fig. 4. Series of the ARPES spectra obtained on graphene/Ni(111), $H_2O$/graphene/Ni(111), and $NH_3$/graphene/Ni(111) along the $\Gamma$–$K$ direction of the Brillouin zone. The amounts of water and ammonia were estimated as 0.5 of the molecular layer. These data were collected with the photon energy of 75 eV.

**Table 1** The interaction energies ($E_{int}$) and the equilibrium distances ($d_0$) between $H_2O$ and the surface of the freestanding graphene layer as obtained for six selected geometries at DFT level with standard PBE functional and when including dispersion correction (PBE-D2).

| Geometry | PBE [a] | | PBE-D2 | |
|---|---|---|---|---|
| | $d_0$ (Å) | $E_{int}$ (meV) | $d_0$ (Å) | $E_{int}$ (meV) |
| C_DOWN | 4.02 | 19 | 2.60 | 139 |
| C_UP | 3.69 | 20 | 3.07 | 83 |
| B_DOWN | 4.05 | 18 | 2.67 | 129 |
| B_UP | 3.70 | 18 | 3.17 | 77 |
| T_DOWN | 4.05 | 19 | 2.64 | 127 |
| T_UP | 3.70 | 19 | 3.18 | 75 |

[a] Data taken from Ref. [15].

**Table 2** The interaction energies ($E_{int}$) and the equilibrium distances ($d_0$) between $H_2O$ ($NH_3$) and the graphene/Ni(111) substrate as obtained for eight selected geometries at PBE-D2 level of theory.

| System<br>Geometry | $H_2O$/graphene/Ni(111) | | $NH_3$/graphene/Ni(111) | |
|---|---|---|---|---|
| | $d_0$ (Å) | $E_{int}$ (meV) | $d_0$ (Å) | $E_{int}$ (meV) |
| C_DOWN | 2.55 | 123 | 3.19 | 127 |
| C_UP | 3.03 | 64 | 2.93 | 143 |
| B_DOWN | 2.64 | 111 | 3.21 | 124 |
| B_UP | 3.11 | 58 | 2.95 | 141 |
| T(C1)_DOWN | 2.63 | 110 | 3.12 | 123 |
| T(C1)_UP | 3.14 | 56 | 2.89 | 148 |
| T(C2)_DOWN | 2.62 | 111 | 3.12 | 125 |
| T(C2)_UP | 3.13 | 58 | 2.91 | 146 |